\documentclass[12pt]{iopart}
\usepackage{epsfig}
\epsfclipon
\newcommand{\nqmean}[1]{\langle\!\langle #1\rangle\!\rangle_q}
\newcommand{\parfrac}[2]{\frac{\partial #1}{\partial #2}}

\newcommand{\tfrac}[2]{\frac{\rmd #1}{\rmd #2}}
\begin{document}
%
%
\title[Physical temperature and the meaning of the $q$ parameter in Tsallis statistics]{Physical temperature and the meaning of the $q$ parameter in Tsallis statistics}
\author{E Ruthotto}
\address{Institut f\"ur Theoretische Physik, Westf\"alische Wilhelms--Universit\"at M\"unster, Wilhelm--Klemm--Str. 9, 48149 M\"unster, Germany}
\ead{eicke.ruthotto@uni-muenster.de}
%
%
\begin{abstract}
We show that the function $\beta(E)$ derived from the density of states of a constant heat capacity reservoir coupled to some system of interest is not identical to the physically measurable (transitive) temperature. There are, however, connections between the two quantities as well as with the Tsallis parameter $q$. We exemplify these connections using the one--dimensional Ising model in the ``dynamical ensemble''.
\end{abstract}
\submitto{\JPA}
%
%
\vspace*{5mm}
There is an ever growing list of systems that are shown---or believed---to be inadequately described by standard Boltzmann--Gibbs statistical
mechanics. All of these systems in some way involve one or more of the
following common features: (i) long--range interactions, (ii) long--time
memory, (iii) (multi--)fractal space--time. In order to tackle these
systems mathematically, Tsallis \cite{tsallis1} introduced a
one--parameter generalization of Shannon's information entropy and, along
the lines of Jaynes \cite{jaynes}, established the framework of generalized
statistical mechanics (GSM) \cite{tsallis2,tmp}.\\
GSM has been successfully applied to a wealth of problems---for a regularly
updated list see \cite{web}. There is, however, to this day some confusion
about the appropriate definition of the thermodynamic temperature.
Furthermore, the physical interpretation of the nonextensivity parameter $q$
is---in most cases---still an open question.\\
This paper is organized as follows: In section 1 we shall give a short introduction to Tsallis entropy, generalized mean values and canonical probability distributions. In section 2 we shall introduce the ``physical temperature'' and the possible transformation to an extensive entropy. We shall show that this transformation is of little consequence for practical calculations. In section 3 we summarize how the ``Tsallis factor'' can be obtained from a microcanonical basis in the case of a ``constant heat capacity'' reservoir, and section 4 is devoted to an application of Tsallis statistics to the nearest--neighbour Ising model in contact with an ideal gas. Our conclusions will be given in section 5.
%
%
\section{Tsallis statistics}
The generalized entropy introduced by Tsallis (and previously in different
contexts by Havrda, Charvat and Daroczy) reads
\begin{equation}
S_q = k_\mathrm{T}\,\frac{1-\sum_{i=1}^\Omega p_i^q}{q-1}\,,\label{tsallisentropy}
\end{equation}
where $i=1,\ldots,\Omega$ labels the possible microstates of the system under consideration, $\left\{ p_i\right\}$ are the microstate
probabilities, $q$ is a real parameter and $k_\mathrm{T}$ is a positive real
constant that approaches Boltzmann's constant $k_\mathrm{B}$ as $S_q$ approaches the Shannon entropy in the limit $q\to
1$:
\begin{equation}
S_q\stackrel{q\to 1}{\longrightarrow} S_1 \equiv S=-k_\mathrm{B}\sum_{i=1}^\Omega p_i\,\ln p_i\,.
\end{equation}
In the following
$k_\mathrm{T}$ will be set to unity.\\
The parameter $q$ determines the degree of nonextensivity (nonadditivity) of
$S_q$ in the following sense: If two statistically independent systems $A$ and
$B$ with probabilities $\left\{ p_i^A\right\}$ and $\left\{p_j^B\right\}$,
respectively, are combined to form the system $A\cup B$, the composite
entropy $S_q^{A\cup B}$ satisfies the relation
\begin{equation}
S_q^{A\cup B} = S_q^A + S_q^B +(1-q)\,S_q^A S_q^B\,.\label{pseudoadd}
\end{equation}
Thus, the entropy is superextensive (superadditive) for $q<1$, extensive
(additive) for $q=1$ and subextensive (subadditive) for $q>1$.\\
$S_q$ has been shown to be positive for all $q$, to exhibit definite curvature
for $q<0$ and $q>0$ (it is constant for $q=0$), and to attain its maximum value
in the case of equiprobability.\\
The connection to thermodynamics has been established by generalizing the
canonical ensemble to the case $q\neq1$, i.e. maximizing the entropy under
the constraints of (i) normalization of the probability distribution and
(ii) knowledge of the energy expectation value.\\
The normalized $q$--expectation value \cite{tmp} of some observable $O$ is
defined as
\begin{equation}
\nqmean{O} = \frac{\sum_{i=1}^\Omega p_i^q O_i}{\sum_{j=1}^\Omega p_j^q}\,,
\end{equation}
where $\left\{O_i\right\}$ are the microstate values of $O$.\\
Maximization of the functional
\begin{equation}
F\left[\left\{ p_i\right\}\right] = S_q - \alpha \left(\sum_{i=1}^\Omega p_i-1\right) -\beta^\mathrm{TMP}\left(\frac{\sum_{i=1}^\Omega p_i^q E_i}{\sum_{j=1}^\Omega p_j^q} -\nqmean{E}\right)
\end{equation}
leads to the generalized canonical probability distribution
\begin{equation}
p_i=\frac{1}{Z_q} \left[ 1-(1-q)\frac{\beta^\mathrm{TMP}}{\sum_{j=1}^\Omega p_j^q} \left( E_i-\nqmean{E}\right)\right]^\frac{1}{1-q}_+\qquad (i=1,\ldots,\Omega)\,,\label{p_tmp}
\end{equation}
where TMP stands for Tsallis--Mendes--Plastino to distinguish the parameter $\beta$ from other choices (introduced below), $Z_q$ is the generalized partition function,
\begin{equation}
Z_q = \sum_{i=1}^\Omega \left[ 1-(1-q)\frac{\beta^\mathrm{TMP}}{\sum_{j=1}^\Omega p_j^q} \left(E_i-\nqmean{E}\right)\right]^\frac{1}{1-q}_+,
\end{equation}
and $[\ldots]_+$ symbolizes the Tsallis cut--off condition, i.e. $[\ldots]_+ = [\ldots]\Theta(\ldots)$, where $\Theta(x)$ is Heaviside's unit step function.\\
With this formalism, the Legendre transform structure of thermodynamics is preserved \cite{yamano}, in particular
\begin{equation}
\parfrac{S_q}{\nqmean{E}} = \beta^\mathrm{TMP}.\label{microtemp}
\end{equation}
In order to solve the manifestly self--referential equations (\ref{p_tmp}) for the probabilities $p_i$, Tsallis et al. introduced the so--called $\beta$--$\beta'$ transformation \cite{tmp}: Using an auxiliary parameter $\beta'$, equation (\ref{p_tmp}) can be rewritten in the form
\begin{equation}
p_i = \frac{ \left[ 1-(1-q)\beta' E_i\right]^\frac{1}{1-q}_+}{\sum_{j=1}^\Omega \left[ 1-(1-q)\beta' E_j\right]^\frac{1}{1-q}_+}\,,\label{beta_prime}
\end{equation}
where 
\begin{equation}
\beta'\left(\beta^\mathrm{TMP}\right) = \frac{\beta^\mathrm{TMP}}{\sum_{j=1}^\Omega p_j^q + (1-q)\beta^\mathrm{TMP}\nqmean{E}}\,.
\end{equation}
Mart\'inez et al. \cite{olm} showed that with equations (\ref{p_tmp}) for the microstate probabilities a maximum of $S_q$ cannot be guaranteed since the concomitant Hessian is not diagonal. To overcome this problem, they suggest a reformulation of the internal energy constraint, thus maximizing the functional
\begin{equation}
F'\left[\left\{ p_i\right\}\right] = S_q - \alpha\left(\sum_{i=1}^\Omega p_i -1\right)-\beta^\mathrm{OLM}\sum_{i=1}^\Omega p_i^q\left(E_i-\nqmean{E}\right)\,.
\end{equation}
Here OLM stands for ``Optimal Lagrange Multiplier''. This procedure leads to the entropy maximizing probabilities
\begin{equation}
p_i = \frac{1}{Z_q^\mathrm{OLM}}\left[ 1-(1-q)\beta^\mathrm{OLM}\left(E_i-\nqmean{E}\right)\right]^\frac{1}{1-q}_+\,,\label{p_olm}
\end{equation}
which can also be written in the form (\ref{beta_prime}) with
\begin{equation}
\beta'\left(\beta^\mathrm{OLM}\right) = \frac{\beta^\mathrm{OLM}}{1+(1-q)\beta^\mathrm{OLM}\nqmean{E}}\,.\label{beta_olm_prime}
\end{equation}
In this case the Legendre structure is altered, and equation (\ref{microtemp}) reads
\begin{equation}
  \parfrac{S_q}{\nqmean{E}} = \beta^\mathrm{OLM}\sum_{j=1}^\Omega p_j^q = \beta^\mathrm{OLM}\left(Z_q^\mathrm{OLM}\right)^{1-q}\,.
\end{equation}
Various other choices of the internal energy constraint have been made, among those the reformulation of the variational problem in terms of the so--called ``escort probabilities'' $P_i=p_i/\sum_j p_j^q$ \cite{tmp, abe1}. Most interesting, however, is a work by Suyari \cite{suyari}, who showed that, if one wants to interpret $S_q$ as the average of some nonextensive information measure $I_q$ and postulates that $I_q$ and $S_q$ satisfy the same nonextensivity relation, one not only has to use normalized $q$--expectation values, but also the ``normalized'' $q$--entropy
\begin{equation}
  S_q^\mathrm{norm} = \frac{1-\sum_{i=1}^\Omega p_i^q}{(q-1)\sum_{j=1}^\Omega p_j^q}\,.
\end{equation}
The same ``normalization'' of the Tsallis entropy has previously been suggested by Rajagopal and Abe \cite{raja} on grounds of form invariance considerations of the generalized Kullback--Leibler relative entropy.\\ 
As will be shown later in this paper, the actual choice of the form of the energy constraint is insubstantial as far as one (i) uses normalized $q$--expectation values and (ii) is only interested in the behaviour of the system as a function of the ``physical temperature'' to be introduced in the following section.
%
%
\section{Physical temperature}
Consider an isolated system $G$ composed of two subsystems $A$ and $B$ in ``weak'' thermal contact. ``Weak'' contact means that the interaction energy between the two systems shall be negligible, thus
\begin{equation}
E_G = E_A + E_B = \mathrm{const.}
\end{equation}
If $A$ and $B$ do not have the same temperature from the start, after some (possibly very long) time, thermal equilibrium will be reached. This equilibrium state is characterized by a maximum of the composite entropy under the constraint of constant total energy. Since the entropy $S_q^{A\cup B}$  satisfies equation (\ref{pseudoadd}), one gets \cite{abe2}
\begin{eqnarray}
0 = \delta S_q^{A\cup B} &=& \left[1-(1-q)S_q^B\right]\parfrac{S_q^A}{\nqmean{E_A}}\,\delta\nqmean{E_A}\nonumber\\
& + &\left[1-(1-q)S_q^A\right]\parfrac{S_q^B}{\nqmean{E_B}}\,\delta\nqmean{E_B}\,,
\end{eqnarray}
which, because of $\delta\nqmean{E_A} = -\delta\nqmean{E_B}$, leads to the equilibrium parameter (``physical temperature'')
\begin{equation}
\frac{1}{1+(1-q)S_q^A}\left(\parfrac{S_q^A}{\nqmean{E_A}}\right) = \frac{1}{1+(1-q)S_q^B}\left(\parfrac{S_q^B}{\nqmean{E_B}}\right) \equiv \beta_\mathrm{phys}\,.\label{betaphys}
\end{equation}
This result is sometimes referred to as the ``generalized zeroth law of thermodynamics''.
For the two cases from the previous section this means that
\begin{equation}
  \beta_\mathrm{phys} = \frac{\beta^\mathrm{TMP}}{\sum_{j=1}^\Omega p_j^q} = \beta^\mathrm{OLM}\,.\label{beta_phys}
\end{equation}
The first identity in (\ref{beta_phys}) has also been obtained by Kalyana Rama \cite{rama} on grounds of the equivalence between canonical and microcanonical ensemble.\\
The necessity of the transformation (\ref{betaphys}) has also been shown by Vives et al. \cite{vives_planes} from quite a different point of view: Using equation (\ref{pseudoadd}) and scaling of the entropy, they established a generalization of the Gibbs--Duhem relation, showing that the Lagrange parameters (in the TMP formalism) are not intensive quantities. By introducing the above transformation of the Lagrange parameters together with a transformation of the entropy,
\begin{equation}
\widehat{S}_q \equiv \frac{\ln\left[1+(1-q)S_q\right]}{1-q}\,,\label{renyi}
\end{equation}
the standard Gibbs--Duhem relation can be reobtained.\\
The same transformations have been deduced by Johal \cite{johal} from the existence of thermal equilibrium of two systems in weak contact and the existence of a Legendre transform structure. The new entropy $\widehat{S}_q$ is an extensive quantity, i.e. $\widehat{S}_q^{A\cup B} = \widehat{S}_q^A + \widehat{S}_q^B$, and by inserting equation (\ref{tsallisentropy}) one can see that $\widehat{S}_q$ is the Renyi entropy:
\begin{equation}
\widehat{S}_q = S_\mathrm{R} \equiv \frac{\ln\sum_{i=1}^\Omega p_i^q}{1-q}\,.
\end{equation}
As has been shown by Lenzi et al. \cite{lenzi}, extremization of $S_\mathrm{R}$ with the constraints of normalization and normalized $q$--expectation value of the energy leads to the OLM distribution (\ref{p_olm}) with 
\begin{equation}
\beta = \beta_\mathrm{phys} = \parfrac{S_\mathrm{R}}{\nqmean{E}}\,.
\end{equation}
The complete Legendre transform structure is preserved \cite{lenzi, abe3}.\\
Now, it is interesting to note that for actual calculations the probabilities (\ref{p_tmp}) or (\ref{p_olm}) must be determined either iteratively or by means of the $\beta$--$\beta'$ transformation. The latter procedure entails calculation of the probabilities in terms of the auxiliary parameter $\beta'$, computation of the relevant averages and subsequent determination of the associated temperature \cite{tmp, penna}. If one is interested in the behaviour of the system under consideration with the ``physical temperature'', one gets for all formulations (TMP, OLM, normalized entropy, Renyi entropy) the same relation:
\begin{equation}
\beta_\mathrm{phys} = \frac{\beta'}{1-(1-q)\beta'\nqmean{E}}\,.\label{common_beta}
\end{equation}
However, if one wishes to address the subject of thermodynamic stability, some subtleties have to be considered. The generalized free energy that is minimized by the OLM distribution (\ref{p_olm}) is associated by Legendre transformation not with the Tsallis entropy, but with the Renyi entropy \cite{abe3}. It reads
\begin{equation}
F_q^\mathrm{OLM} = \nqmean{E}-\frac{1}{\beta_\mathrm{phys}}S_\mathrm{R} = \nqmean{E}-\frac{1}{\beta_\mathrm{phys}}\ln Z_q^\mathrm{OLM}\,.\label{fq}
\end{equation}
Defining
\begin{equation}
\ln \widehat{Z}_q^\mathrm{OLM} = \ln Z_q^\mathrm{OLM}-\beta_\mathrm{phys}\nqmean{E}\,,
\end{equation}
thermodynamic relations recover their familiar forms \cite{lenzi}:
\begin{equation}
F_q^\mathrm{OLM} = -\frac{1}{\beta_\mathrm{phys}}\ln \widehat{Z}_q^\mathrm{OLM}\,,
\end{equation}
\begin{equation}
\nqmean{E} = -\parfrac{}{\beta_\mathrm{phys}} \ln \widehat{Z}_q^\mathrm{OLM}\,,
\end{equation}
and
\begin{equation}
C_q^\mathrm{OLM} = \parfrac{\nqmean{E}}{T_\mathrm{phys}} = -T_\mathrm{phys}\frac{\partial^2F_q^\mathrm{OLM}}{\partial T_\mathrm{phys}^2}\,.\label{specific_heat}
\end{equation}
Since thermodynamic stability is commonly associated with positivity of the specific heat, one gathers from equation (\ref{specific_heat}) that $F_q^\mathrm{OLM}$ is required to be a concave function of the physical temperature. As will be illustrated in section 4, due to the $\beta_\mathrm{phys}$--$\beta'$ transformation, this criterion is not always fulfilled, and some kind of Maxwell construction must be applied. Also, for reasons of concavity of the Renyi entropy $S_\mathrm{R}$, the parameter $q$ is restricted to the interval $0<q\leq1$.
%
%
\section{Microcanonical derivation of the Tsallis factor}
As in the previous section we consider an isolated system $G$ composed of two weakly interacting systems $S$ (``System'') and $B$ (``Bath''). Once again, the energies satisfy the relation
\begin{equation}
E_G = E_S + E_B = \mathrm{const.}
\end{equation}
The principle of equal a--priori probabilities demands that, in a situation where the total energy is $E_G$ and the system energy is $E_S$ (the bath energy is accordingly $E_G-E_S$), all compatible states are equally probable, thus
\begin{equation}
p_S(E_S) = \frac{\Omega_S(E_S)\Omega_B(E_G-E_S)}{\Omega_G(E_G)}\,,\label{microprobs}
\end{equation}
where $\Omega_X(E)$ is the number of states (the structure function) of system $X$  with energy $E$, and $\Omega_G(E_G)$ is given by
\begin{equation}
\Omega_G(E_G) = \int_{0}^{E_G} \Omega_S(E_S)\Omega_B(E_G-E_S)\,\rmd E_S\,.\label{omegatotal}
\end{equation}
From the definition
\begin{equation}
\beta(E) = \tfrac{}{E}\,\ln \Omega_B(E) \label{beta_e}
\end{equation}
the Boltzmann factor 
\begin{equation}
p_S(E_S) \propto \exp(-\beta E_S)
\end{equation}
can be obtained, provided one of the following conditions holds:
\begin{itemize}
\item The temperature of the bath is exactly constant (``constant $T$'' derivation) or
\item The energy of the bath is large compared to the system energy: $E_S\ll E_B$ (``small $E_S$'' derivation)
\end{itemize}
The latter condition, however, does not unambiguously lead to Boltzmann--Gibbs statistics \cite{tatsuaki}. Indeed, every distribution of the form
\begin{equation}
p_S(E_S) \propto \exp_\mathcal{Q}(-\beta E_S)\label{tsallis_factor}
\end{equation}
can be obtained, where $\exp_\mathcal{Q}(x)=(1+\mathcal{Q}x)^{1/\mathcal{Q}}$ is a $\mathcal{Q}$--deformed exponential that converges to the ordinary exponential in the limit $\mathcal{Q}\to 0$. Of this, the ``Tsallis factor'' is but a special case (for $\mathcal{Q}=1-q$).\\
By an extension of the ``small $E_S$'' derivation to the case where $E_S$ is not necessarily small (i.e. there is no truncation of some series expansion involved), Almeida \cite{almeida1, almeida2} has derived the above distribution (\ref{tsallis_factor}) under the assumption that the heat capacity of the reservoir is exactly constant:
\begin{equation}
C_B = \tfrac{E_B}{T} = \frac{1}{\mathcal{Q}} = \mathrm{const.}\label{const_heat}\qquad\Leftrightarrow\qquad \tfrac{}{E_B}\left(\frac{1}{\beta(E_B)}\right) = \mathcal{Q}=\mathrm{const.}
\end{equation}
This includes the true heat reservoir limit, where the temperature is exactly constant, irrespective of the amount of heat gained or lost by the bath. In this case the heat capacity of the bath is infinite, thus $\mathcal{Q}=0$, which is the Boltzmann case. Note that with (\ref{const_heat}) the parameter $\mathcal{Q}$ and therefore the Tsallis parameter $q$ is linked to a physical property of the heat bath and thus gets a physical interpretation. So $q$ is restricted to $q\leq1$, if one does not allow for negative heat capacities.\\
Starting from equation (\ref{const_heat}) together with (\ref{beta_e}), Tatsuaki \cite{tatsuaki} established the following differential equation for the structure function $\Omega_B(E_B)$ of the bath:
\begin{equation}
\rmd\ln\Omega_B(E_B) = \frac{\beta_0}{1+\mathcal{Q}\beta_0 E_B}\,\rmd E_B\,,
\end{equation}
where $\beta_0 = \beta(0)$ is an integration constant. Integrating this equation, one finds the solution
\begin{equation}
\Omega_B(E_B) = \Omega_B(0)\,\exp_\mathcal{Q}(\beta_0 E_B)\,.
\end{equation}
From this and equation (\ref{beta_e}) one gets for the $\beta$ function of the constant heat capacity environment
\begin{equation}
\beta(E_B) = \frac{\beta_0}{1+\mathcal{Q}\beta_0 E_B}\,. \label{beta_null}
\end{equation}
Using $E_B=E_G-E_S$, one obtains 
\begin{equation}
\Omega_B(E_G-E_S) = \Omega_B(0)\,\exp_\mathcal{Q}\left[\beta_0 E_G\right]\,\exp_\mathcal{Q}\left[-\beta(E_G)E_S\right]\,,
\end{equation}
and, inserting this into equation (\ref{microprobs}) and using (\ref{omegatotal}), one finally arrives at
\begin{equation}
p_S(E_S) = \frac{\Omega_S(E_S)\exp_\mathcal{Q}\left[-\beta(E_G)E_S\right]}{\int_0^{E_G}\Omega_S(E_S)\exp_\mathcal{Q}\left[-\beta(E_G)E_S\right]\,\rmd E_S}\,,\label{distribution}
\end{equation}
with 
\begin{equation}
\beta(E_G) = \frac{\beta_0}{1+\mathcal{Q}\beta_0 E_G}\,.
\end{equation}
Comparison with equation (\ref{beta_prime}) shows that the two probability distributions are the same with
\begin{equation}
\beta(E_G) = \beta'\,,
\end{equation}
and from equation (\ref{common_beta}) one gets that the (inverse) physical temperature is given by
\begin{equation}
\beta_\mathrm{phys} = \beta\left(\nqmean{E_B}\right) = \frac{\beta_0}{1+\mathcal{Q}\beta_0\left(E_G-\nqmean{E_S}\right)}\,,
\end{equation}
where $\nqmean{E_B}=\nqmean{E_G-E_S}=E_G-\nqmean{E_S}$ has been used.\\
For the sake of completeness, let us mention that already in 1994 Plastino and Plastino \cite{plastino} showed that for a system in contact with a {\it finite} heat bath with structure function $\Omega(E)\propto E^\alpha$ the resulting equilibrium distribution is of Tsallis form with $q=(\alpha-1)/\alpha$.
%
%
\section{The Ising model in the ``dynamical ensemble''}
As an application, let us now consider a one--dimensional Ising model ($S$) of $N$ spins with periodic boundary conditions coupled to an ideal gas ($B$) with $M$ degrees of freedom \cite{gerling}.\\
The Ising model is given by the Hamiltonian
\begin{equation}
H = \sum_{i=1}^N(1-\sigma_i\sigma_{i+1})\,,
\end{equation}
where $\sigma_i$ is a spin variable with values $\pm 1$, corresponding to the spin pointing ``up'' or ``down''. The total energy of the composite system is given by
\begin{equation}
E_G = E_S+ E_B =N\varepsilon + Mk = \mathrm{const.}\,,
\end{equation}
where $\varepsilon=E_S/N$ is the energy per spin of the Ising system, and $k=E_B/M$ is the kinetic energy per degree of freedom of the gas.\\
With the known structure function of an ideal gas,
\begin{equation}
\Omega_B(E_B) = C\, E_B^\frac{M-2}{2}\,,
\end{equation}
$C$ being some constant and
\begin{equation}
E_B = E_G-E_S\,,
\end{equation}
we get for the (unnormalized) probability of finding the composite system $G$ in a state where $S$ has the energy $E_S$:
\begin{equation}
p_S(E_S) \propto \Omega_S(E_S)\left(E_G-E_S\right)^\frac{M-2}{2}.\label{p_G}
\end{equation}
Since the ideal gas is a constant heat capacity environment,
\begin{equation}
C_M = \frac{M-2}{2} \equiv \frac{1}{\mathcal{Q}}\,,
\end{equation}
we expect the equilibrium distribution of the Ising system to be of Tsallis form with
\begin{equation}
q = 1-\mathcal{Q} = 1-\frac{2}{M-2}\,.\label{q}
\end{equation}
From equation (\ref{beta_e}) follows
\begin{equation}
\beta(E_B) = \frac{M-2}{2E_B} = \frac{1}{2k}\left(1-\frac{2}{M}\right)\,.\label{beta_gas}
\end{equation}
One sees that in the limit of an infinite heat bath $M\to\infty$ the statement of the equipartition theorem is recovered. In that limit, $\beta(E_B)$ is the (inverse) physical temperature of the bath.\\
To be able to calculate the energy of the Ising system as a function of the physical temperature by means of Monte Carlo simulations, it is necessary to determine the normalized $q$--expectation value of the system energy $\nqmean{E_S}$ at given total energy $E_G$. This is accomplished using the generalized Metropolis sampling algorithm by Andricioaei and Straub \cite{as1,as2}. Weighting the microstate energies $E_S^{(i)}$ with $p_S^q\left(E_S^{(i)}\right)$, the transition probabilites from state $i$ to $j$, related by single spin flips, become
\begin{equation}
P(i\to j) = \min\left[1, \left(\frac{E_G-E_S^{(j)}}{E_G-E_S^{(i)}}\right)^\frac{M-4}{2}\right]\,,
\end{equation}
where equations (\ref{p_G}) and (\ref{q}) have been used. From equations (\ref{common_beta}) and (\ref{beta_gas}) we obtain the common physical temperature of the bath and the Ising system:
\begin{equation}
T_\mathrm{phys} = \frac{2}{M-2}\left(E_G-\nqmean{E_S}\right)\,.
\end{equation}
Figure 1 shows the energy per spin of an $N=128$ Ising system as a function of the physical temperature for different sizes of the gas reservoir. The data points were obtained by means of Monte Carlo simulations, while the lines represent exact results, calculated from equations (\ref{beta_prime}) and (\ref{common_beta}) using the known structure function of the Ising model and the appropriate $q$ values given by (\ref{q}). The curve for $q=1$ is also given for comparison.
\begin{center}
\begin{figure}[h!]
\centerline{{\epsfig{file=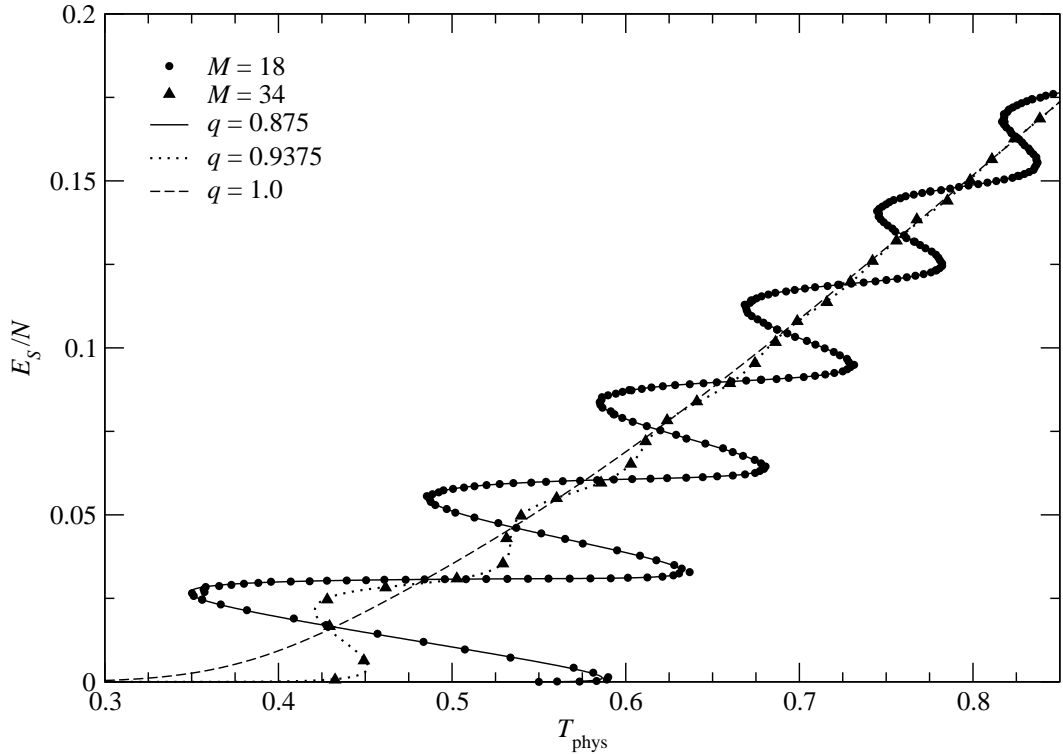, width=14cm,  angle=0}}}
\caption{Specific energy of an $N=128$ Ising system for bath sizes $M=18$ and $M=34$ as a function of the physical temperature. Symbols represent simulation results, lines are exact results obtained from the $q$--canonical distribution for the appropriate $q$ values.}
\end{figure}
\end{center}
Besides the fact that the simulation and the exact results are in good agreement, one finds that the energy is a multi--valued function of the physical temperature. Thus, we take a look at the generalized free energy defined by (\ref{fq}) and discover a behaviour that is similar to what was described by Lima et al. in \cite{penna} for the TMP case. Figure 2 shows the free energy (a) and the internal energy (b) for the case $q=\frac{10}{11}\ $ ($M=24$). The solid lines represent $F_q^\mathrm{OLM}$ and $\nqmean{\varepsilon}$ calculated by means of the $\beta_\mathrm{phys}$--$\beta'$ transformation, while the broken curves were calculated from the iteratively determined probabilities $\left\{p^\mathrm{OLM}\right\}$. The $\beta_\mathrm{phys}$--$\beta'$ transformation produces a closed loop in the free energy curve, whereas the iterative method reveals metastable branches. The behaviour depends on whether one starts from low or high temperatures. By always choosing the branch with the lowest free energy the metastable states are eliminated, leading to a discontinuous internal energy as indicated in figure 2 (b).\\
\begin{center}
\begin{figure}[h!]
\centerline{{\epsfig{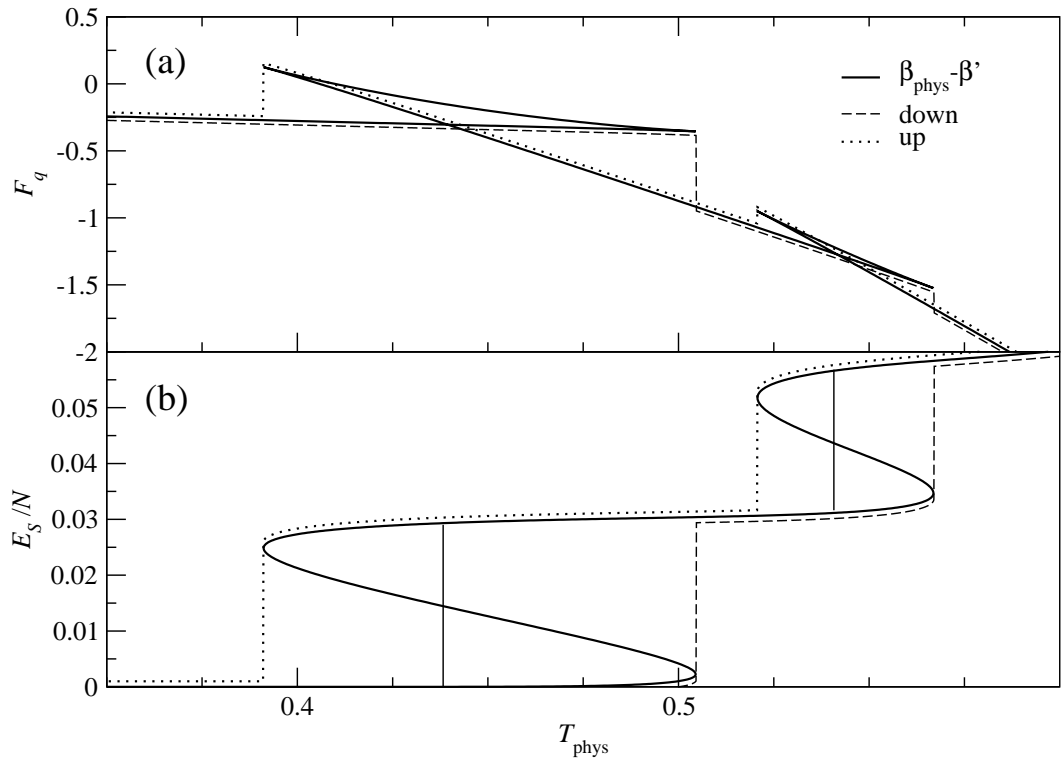}}}
\caption{Free energy (a) and internal energy (b) of an $N=128$ Ising system for $q=\frac{10}{11}$. The solid lines are obtained via the $\beta_\mathrm{phys}$--$\beta'$ transformation, the broken curves are calculated iteratively. Latter curves have been slightly displaced for better visualization.}
\end{figure}
\end{center}
%
%
\section{Conclusions}
We have shown that the function $\beta(E_B)$, defined in analogy to the Boltzmann--Gibbs ($q=1$) microcanonical ensemble as the logarithmic derivative of the structure function of a constant heat capacity reservoir $B$ coupled to some system $S$, does not describe the common physical temperature of the two systems. Instead, $\beta(E_B)$ at $E_B=E_G$, where $E_G$ is the constant total energy of the system+bath, corresponds to the auxiliary parameter $\beta'$ introduced in Tsallis canonical ensemble theory. From this the physical temperature can be obtained by a $\beta_\mathrm{phys}$--$\beta'$ transformation (which is the same for {\it all} formulations of the generalized canonical ensemble), provided one knows the normalized $q$--expectation value of the bath energy $E_B$ at constant $E_G$.\\
Using the one--dimensional Ising model in contact with a finite ideal gas reservoir as an example, we have shown that this transformation can be performed adopting the generalized Metropolis Monte Carlo sampling algorithm of Andricioaei and Straub. The curves of the free and internal energy thus obtained, show the same ``reentrant behaviour'' already known from the TMP canonical ensemble. Choosing always the states with the lowest free energy, single--valued curves can be obtained. The curves of the internal energy become discontinuous, while the free energies have discontinuous first derivatives.\\
In summary we can say that a physical system that is in weak contact with a constant heat capacity reservoir is described by a Tsallis canonical distribution with an index $q$ that is determined by the nature of the bath (including the case of an infinite heat capacity, in which the Boltzmann--Gibbs distribution is recovered). The factor $\beta'$ in the $q$--exponential is only the inverse temperature of the system if $q=1$. 
%
%
\ack The author wishes to thank Prof. Dr. G. M\"unster for the support and fruitful discussions during the preparation of this paper.
%
%
\section*{}

\end{document}